\documentclass{article}
\usepackage{spconf,amsmath,graphicx}
\usepackage{amssymb,amsmath,bm}
\usepackage{booktabs}

\title{The AS-NU System for the M2VoC Challenge}
\name{\begin{tabular}{c}
    Cheng-Hung Hu$^1$, Yi-Chiao Wu$^2$, Wen-Chin Huang$^{1,2}$, Yu-Huai Peng$^1$, Yu-Wen Chen$^1$, Pin-Jui Ku$^1$,\\
    Tomoki Toda$^2$, Yu Tsao$^1$, Hsin-Min Wang$^1$
	\end{tabular}
}
\address{$^1$Academia Sinica, Taiwan
		$^2$Nagoya University, Japan}

\begin{document}
\ninept
\maketitle
\begin{abstract}
This paper describes the AS-NU systems for two tracks in Multi-Speaker Multi-Style Voice Cloning Challenge (M2VoC). The first track focuses on using a small number of 100 target utterances for voice cloning, while the second track focuses on using only 5 target utterances for voice cloning. Due to the serious lack of data in the second track, we selected the speaker most similar to the target speaker from the training data of the TTS system, and used the speaker's utterances and the given 5 target utterances to fine-tune our model. The evaluation results show that our systems on the two tracks perform similarly in terms of quality, but there is still a clear gap between the similarity score of the second track and the similarity score of the first track.
\end{abstract}

\section{Introduction}
Speech synthesis refers to the modeling of speech generation by conditioning on certain attributes including text, speaker identity, pitch, etc. Although recent text-to-speech (TTS) \cite{shen2018natural, li2019neural,Ren2019FastSpeechFR} and waveform generation systems \cite{wavenet, samplernn, prenger2019waveglow} have accomplished human-like naturalness, most of them are built upon deep neural network-based models, which are typically trained on a large-scale single-speaker dataset containing tens of hours of speech, so they can only synthesize speech of the training speaker. It is therefore of great interest to learn the voice of an unseen speaker with only a few minutes of speech samples. Such a challenging task is referred to as voice cloning, speaker adaptation or few-shot/one-shot speech synthesis \cite{voice-cloning, jia2018transfer, chen2018sample, nachmani2018fitting, Kons2019, cooper2020zero}.

The multi-speaker multi-style voice cloning (M2VoC) challenge \cite{xie2021M2VoC}  aims to better understand different voice cloning techniques by utilizing a freely-available common dataset to share views about problems and challenges faced by current techniques.
% In addition to learning the speaker identity of the unseen speaker, this challenge also emphasizes the cloning of speaking style, which is less focused in the literature.
The challenge consists of two tracks, where Track~1 is the \textit{few-shot track}, and Track~2 is the \textit{one-shot track}, with 100 and 5 target utterances for cloning, respectively. In this paper, we describe the AS-NU systems for the challenge and present some evaluation results.

\section{System Overview}
\label{sec:system OVerview}
The AS-NU voice cloning systems are built based on Tacotron2 \cite{shen2018natural} and Parallel WaveGAN (PWG) \cite{yamamoto2020parallel}. Tacotron2 predicts the target mel-spectrogram according to the phoneme sequence and the speaker embedding extracted by corresponding pretrained models. Then, the PWG vocoder generates the final synthetic speech based on the predicted mel-spectrogram. In this section, the preprocessing of transcription is first presented. Then, the acoustic feature and the speaker embedding extraction process are described. Finally, we introduce the architectures of the Tacotron2 model and the PWG vocoder and the training and adaptation processes. 

\subsection{Data Preparation}
\subsubsection{Corpus}
Four datasets are provided by the challenge organizer, including, Multi-speaker training speech data (MST), Target speaker validation speech set (TSV), Target speaker testing speech set (TST), and Test text set (TT). MST contains two subsets, namely AIShell3 and Originbeat. AIShell3 contains roughly 85 hours of speech recordings spoken by 218 native Mandarine Chinese speakers while Originbeat contains only a male speaker and a female speaker. We only used AIShell3 to train our multi-speaker TTS system. The AIShell3 corpus contains 63,262 training utterances and 24,773 testing utterances. The TSV set containing four speakers is provided for validation, including two speakers for Track~1 and two speakers for Track~2. We used this dataset to determine our model architectures and hyperparameters. TST contains the target speakers for the two tracks, and there are three speakers for each track. TT contains 300 testing texts for Track~1 and 200 testing texts for Track~2 for generating the cloned voice of the speakers in TST. The sampling rate of the AIShell3 utterances is 44,100~Hz and the bit-depth is 16.  The sampling rate of the TSV and TST utterances is 48,000~Hz and the bit-depth is 32. To unify the audio format, all the 48,000~Hz/32~bit utterances were converted to 44,100~Hz/16~ bit. 

\subsubsection{Text Processing}

The text (Chinese characters) and syllable (Pinyin) transcription of audio files are provided by the organizer of the challenge. To make our systems generate speech with clear pronunciation of each syllable, spaces were inserted between the syllables in the Pinyin transcription. An ending token was appended to the end of each transcription file to stop Tacotron2 from generating subsequent frames. The pretrained G2P Mandarin model provided by Montreal Forced Aligner\cite{mcauliffe2017montreal} was adopted to convert the Pinyin-level transcription into the phone-level transcription. The pretrained G2P model is implemented based on Sigmorphon 2020 G2P task baseline and trained using the Pynini package on the GlobalPhone dataset. 

Because Tacotron2 is an autoregressive (AR) model, it becomes vulnerable to the error accumulation in the AR process when generating long utterances. To tackle this weakness, each test text was divided into multiple sub-texts according to the punctuation marks. Then, the short utterances generated from these sub-texts were concatenated to reassemble the entire output utterance. Although this method may cause unnecessary pauses where the punctuation marks locate, the improvements in quality and prosody prediction are quite significant in our preliminary experiments. Note that this procedure was only used in the inference phase.

\subsubsection{Acoustic Feature and Speaker Embedding}
\label{sec:features}
The mel-spectrogram feature was extracted from an utterance using the ESPNet \cite{espnet} toolkit. The frequency range in mel basis (fmin--fmax), number of mel-filterbanks, FFT window size, and FFT window shift were set to 80--7600~Hz, 80, 2048, and 550, respectively. The speaker embedding of an utterance was extracted following the recipe\footnote{{https://kaldi-asr.org/models/m8}} of SITW X-vector System 1a \cite{snyder2018x}. Each utterance was first downsampled to 16,000~Hz, and then the 512-dimensional utterance-wise x-vector was extracted. The utterance-wise x-vectors corresponding to a speaker were further averaged to form the speaker-wise x-vector, which was used as the speaker embedding required by the Tacotron2 model.

\begin{figure}[t]
  \includegraphics[width=0.45\textwidth]{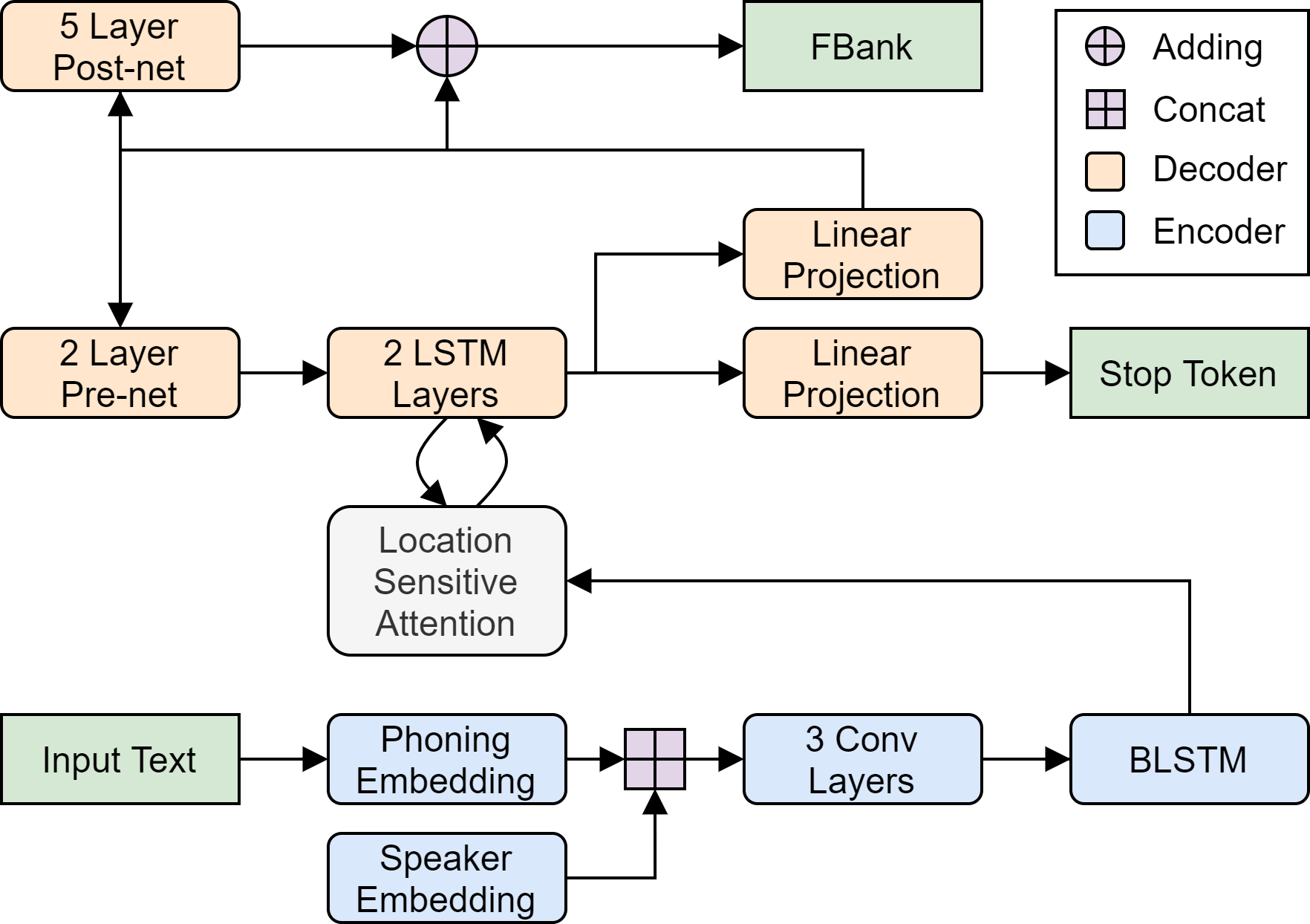}
  \centering
  \caption{
    Multi-speaker Tacotron2 system.
  }
  \label{fig:taco2}
\end{figure}

\subsection{Tacotron2}
As shown in Fig.~\ref{fig:taco2}, the phonetic embedding is extracted based on the input text using a trainable embedding table. Then, the phonetic embedding is concatenated with the corresponding speaker embedding. The Encoder converts the embedding sequence into a representation sequence. The Decoder is an AR model that predicts the next frame of the mel-spectrogram feature based on the mel-spectrogram of previous time steps and the encoded representation sequence. The location sensitive attention (LSA) \cite{LSA} uses cumulative attention weights from previous decoder time steps as an additional feature to alleviate the problem of the Decoder repeating or ignoring subsequences.

The Encoder contains three 1D convolutional layers and a BLSTM layer. The channel size and kernel size of the 1D convolutional layers are set to [512, 512, 512] and [5, 5, 5], respectively, and the kernel size of the BLSTM layer is set to 512. The Decoder consists of the Pre-net, Post-net, two LSTM Layers, and two Linear Projection Modules. The Pre-net contains two Linear Modules with hidden size [512, 512], while the Post-net consists of five 1D convolutional layers with batchnorm, tanh, and Dropout modules. The kernel size of two-layered LSTM is set to [1024, 1024], and the output size of the Linear Projection module is set to [80, 1].

The overall loss is an equal combination of L1, L2, and BCE losses. In the training stage, a speaker-independent (SI) Tacotron2 was first trained with 120 epochs using the AIShell3 training set. After that, two fine-tuning approaches were adopted for Track~1 and Track~2, respectively. For Track~1, the SI model was simply fine-tuned using 95 target utterances with the speaker-wise x-vector while the remaining 5 target utterances were used for validation. For Track~2, to tackle the shortage of the target utterances, based on probabilistic linear discriminant analysis (PLDA) \cite{PLDA}, the speaker-wise x-vector of the 5 target utterances was used to select the speaker most similar to the target speaker from the AIShell3 training set. Then, all the utterances of the selected speaker were mixed with the 5 target utterances to transfer the SI model to the speaker-dependent (SD) model through fine-tuning. In this process, only speakers with more than 100 utterances were considered candidates. Table~\ref{tab:track2 chosen speaker} shows the selected AIShell3 speakers for the Track~2 target speakers. We spot-checked certain utterances of selected speakers and found that the timbre of these selected speakers was indeed close to that of the corresponding target speaker. For both tracks, each SD Tacotron2 model was fine-tuned with 880 epochs.

\begin{table}[]
    \centering
    \begin{tabular}{cc}
        \toprule
        Target Speaker & Most Similar Speaker \\
        \midrule
        Track~2 Speaker~3 & SSB0435 \\
        Track~2 Speaker~4 & SSB1100 \\
        Track~2 Speaker~5 & SSB0139  \\
        \bottomrule
    \end{tabular}
    \caption{The selected speakers for data augmentation.}
    \label{tab:track2 chosen speaker}
\end{table}

\subsection{Parallel WaveGAN}
For the speech synthesis module, PWG~\cite{yamamoto2020parallel} is adopted as the vocoder of our system. The PWG vocoder is a GAN-based generative model, which consists of generator ($G$), discriminator ($D$), and multi-STFT losses modules. 

Specifically, the training loss $ L_{\mathrm{G}}$ of the generator, including a generator adversarial loss ($ L_{\mathrm{adv}}$) and a multi-resolution STFT-based loss ($ L_{\mathrm{sp}}$), is formulated as
\begin{align}
L_{\mathrm{G}}(G, D)=L_{\mathrm{sp}}(G)+\lambda_{\mathrm{adv}} L_{\mathrm{adv}}(G, D),
\label{eq:lg}
\end{align}
where $\lambda_{\mathrm{adv}}$ denotes the weight and is set to 4.0 in our system.
The $ L_{\mathrm{adv}}$ loss is formulated as
\begin{align}
L_{\mathrm{adv}}(G, D)=\mathbb{E}_{\boldsymbol{z} \in N(0, I)}\left[(1-D(G(\boldsymbol{z})))^{2}\right],
\label{eq:ladv}
\end{align}
where $\boldsymbol{z}$ is a Gaussian noise sequence drawn from a zero mean and standard deviation Gaussian distribution, denoted as $N(0, I)$. Note that all auxiliary features of the generator are omitted in this section for simplicity. The $ L_{\mathrm{sp}}$ loss is formulated as 
\begin{align}
L_{\mathrm{sp}}(G)=\frac{1}{M}\sum_{m=1}^{M}(L^{(m)}_{\mathrm{sc}}(G)+L^{(m)}_{\mathrm{mag}}(G)),
\label{eq:lsp}
\end{align}
where $M$ denotes the number of STFT setting groups, and the spectral convergence loss ($L^{(m)}_{\mathrm{sc}}$) and the log STFT magnitude loss ($L^{(m)}_{\mathrm{mag}}$) are calculated on the basis of the STFT features extracted using the settings of the $m$  group. Three STFT setting groups including [1024, 2048, 4096] FFT sizes, [120, 240, 480] frame shifts, and [600, 1200, 2400] window lengths are adopted in our system. More details of $L^{(m)}_{\mathrm{sc}}$ and $L^{(m)}_{\mathrm{mag}}$ can be found in~\cite{yamamoto2020parallel} and the open-source repository\footnote{{https://github.com/kan-bayashi/ParallelWaveGAN}}. Furthermore, the discriminator is trained to minimize the adversarial loss ($ L_{\mathrm{D}}$) formulated as 
\begin{align}
&L_{\mathrm{D}}(G, D) \nonumber \\
&=\mathbb{E}_{\boldsymbol{x} \in p_{\mathrm{data}}}\left[(1-D(\boldsymbol{x}))^{2}\right]+\mathbb{E}_{\boldsymbol{z} \in N(0, I)}\left[D(G(\boldsymbol{z}))^{2}\right],
\label{eq:ld}
\end{align}
where $\boldsymbol{x}$ denotes the natural samples, and $p_{\mathrm{data}}$ denotes the data distribution of the natural samples.

The PWG generator adopts a non-AR WaveNet-like~\cite{wavenet} architecture, which is composed of 30 residual blocks. Each residual block includes a stacked dilated convolution neural network (DCNN), a gated structure, skip and residual connections, and mel-spectrogram as the auxiliary features. The dilation sizes exponentially increase with a base of two and exponents of their layer indexes within every 10 blocks. The PWG discriminator adopts a simpler architecture consisting of only 10 stacked DCNN layers, and the dilation sizes also exponentially increase in the same manner. For both the generator and discriminator, the kernel size is three, and the CNN channel number is 64.

For both M2Voc tracks, a SI PWG vocoder was first built using the AIShell3 training set, and then fine-tuned to the SD PWG vocoder for each target speaker. All parameters of the PWG generator and discriminator were updated during the adaptation phase. For Track~1, each SD PWG vocoder was adapted using the 95 training utterances of the corresponding target speaker. For Track~2, we follow the same process used in our Tacotron2 fine-tuning stage, that is, select utterances in the AIShell3 corpus according to the speaker similarity for speaker adaptation. Specifically, each SD PWG vocoder was adapted using the five target utterances and 90 utterances of the similar speaker. The number of the adaptation iterations was 7000.

\section{Experimental Results}
\label{sec:exp}
In this section, the external subjective evaluation results provided by the M2VoC organizer are presented. Specifically, the M2VoC organizer conducted subjective evaluations online to measure the quality, speaker similarity, and style similarity of the submitted synthetic utterances. All measurements were based on the mean opinion scores (MOS) given by listeners. The range of MOS was 1--5. The higher the score, the better the performance. For the quality evaluation, the listeners were asked to evaluate the naturalness of the synthetic utterances. For the speaker and style similarities, the listeners evaluated the specific similarity of each synthetic utterance and the corresponding two reference utterances.

\begin{figure}[t]
  \includegraphics[width=0.45\textwidth]{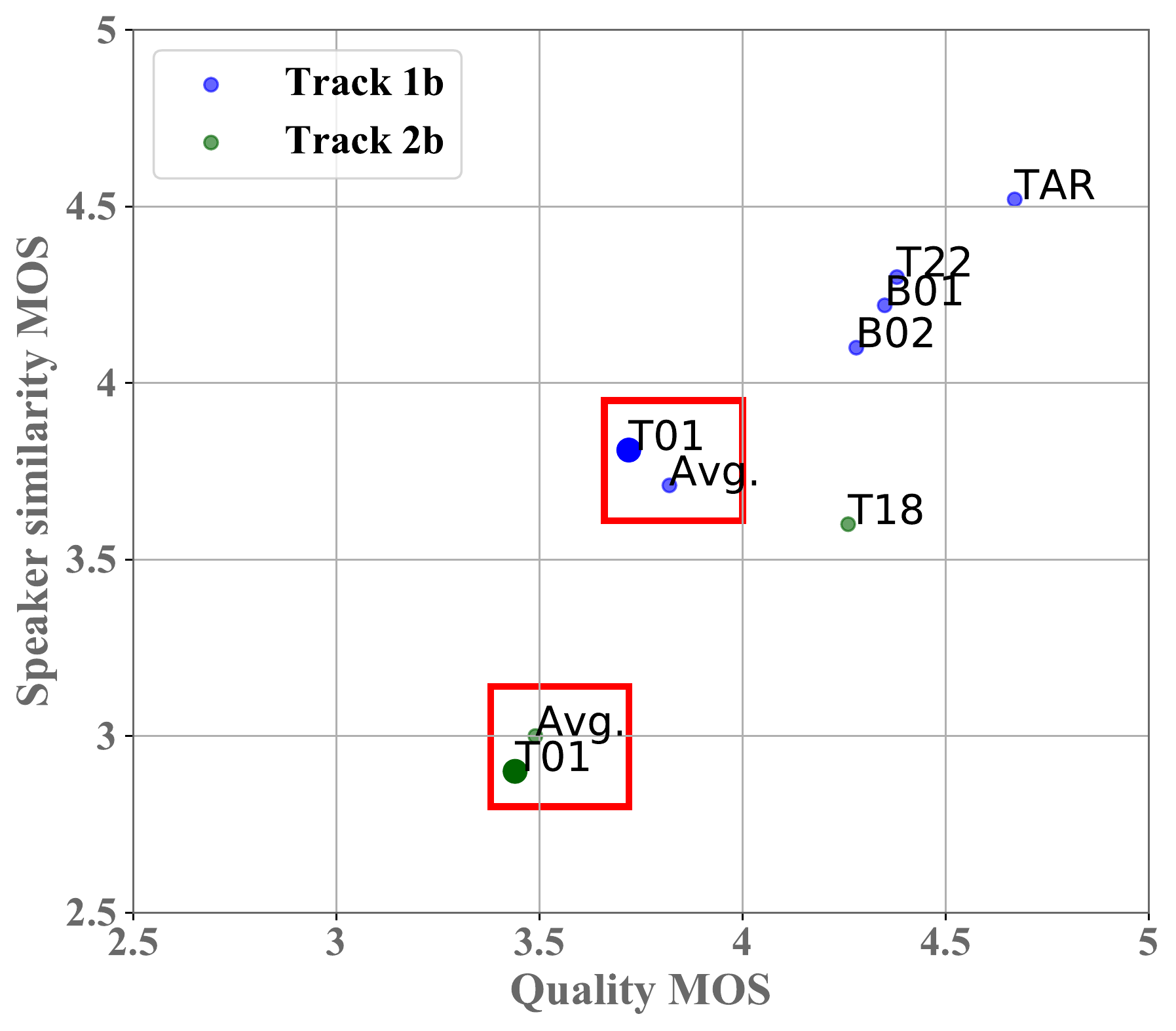}
  \centering
  \caption{
  Quality and speaker similarity MOS results.
  }
  \label{fig:quality v.s. similarity}
\end{figure}

The quality and speaker similarity of both the tracks were evaluated, but the style similarity evaluations were conducted in only the style subset of Track~1. Two rounds of the evaluations were conducted, and only the first round included all submitted systems while the second round included only several top systems. Since our system (T01) was involved in only the first round, Figs.~\ref{fig:quality v.s. similarity} and~\ref{fig:style v.s. similarity} show the results of the first round. For simplicity, only the natural target, official baseline systems, most top system, our system, and overall average results were plotted in these figures. Moreover, since additional corpora were adopted for the extractions of the speaker and phonetic embeddings in our system, our submitted system was evaluated with the systems that also adopted additional corpora (Track~1b and 2b evaluations).

\subsection{Track~1b}
For the Track~1b results, we can find that the performances of our system are close to the average performances. Specifically, we privately plotted the linear regression line of the scores of all submitted systems, and we found that our system achieves higher speaker and style similarities than that of the average results while the qualities of our synthetic utterances are slightly lower. Since we adopt a naïve implementation of the text-to-mel-spectrogram module, the predicted mel-spectrogram still includes many aliasings. The aliased mel-spectrogram usually makes the vocoder generate unnatural sounds and noisy speech, which markedly degrade the naturalness of the synthetic speech.

\begin{figure}[t]
  \includegraphics[width=0.45\textwidth]{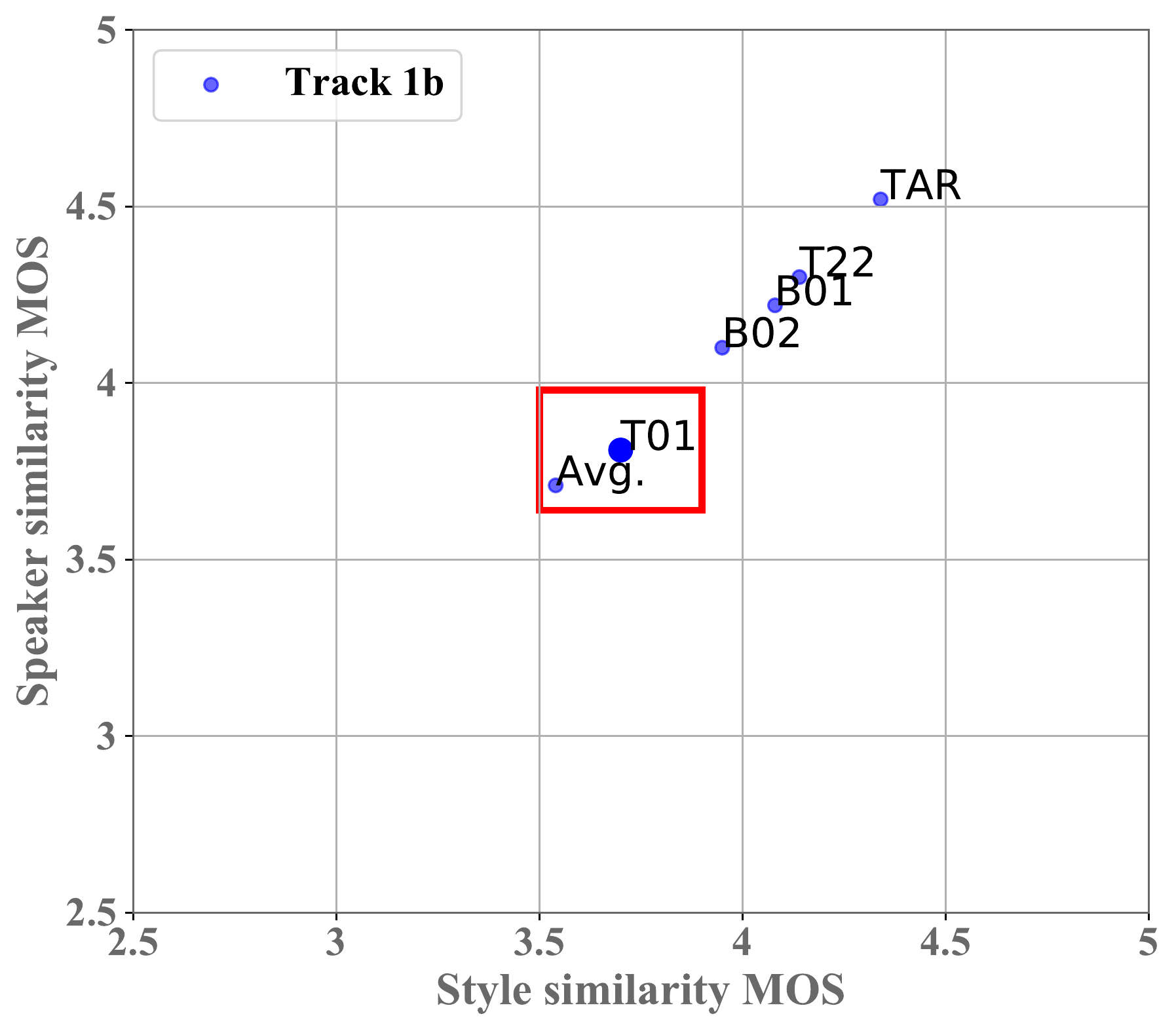}
  \centering
  \caption{
  Style and speaker similarities MOS results.
  }
  \label{fig:style v.s. similarity}
\end{figure}

However, although the speech quality still affects the speaker and style similarities, these similarities are dominated by the timbre and prosody of speech. Therefore, the possible reason for the higher similarity performances compared to the naturalness performance of our system is that the speaker adaptations were applied to both our Tacotron2 and PWG models, which markedly improve the timbre and prosody modeling performances of our system. Furthermore, as shown in Fig.~\ref{fig:style v.s. similarity}, the style scores of all systems are similar to the speaker scores, which indicates that the style similarity is highly related to the speaker similarity. The tendency also confirms our assumption that the style and speaker similarities are more speaker-dependent than the speech naturalness. The results also show the importance of the speaker adaptation.

\subsection{Track~2b}
For the Track~2b results, we can find that our system also achieves similar performances as the overall average results. However, compared to our Track~1b results, the similarity of our synthetic utterances markedly degrades (3.81 to 2.90) while the degradation of the naturalness of our synthetic utterances is less (3.72 to 3.44). Since the available target utterances of Track~2 is much limited, the results also confirm our assumption that speaker similarity is more speaker-dependent than the speech naturalness. Although we already applied speaker-similarity-based data augmentation, advanced data selection may be required for further improving the performance of our system.

\subsection{Discussion}
According to the Track~1b and 2b results, we may conclude that a much larger corpus including more diverse speakers and utterances is required for building SI Tacotron2 and PWG models. That is, although the AIShell3 corpus already included more than 200 speakers and 60,000 utterances for training, the target speaker adaptation was still essential for prosody and timbre modeling of our system. Therefore, there is still room for improving data augmentation methods and modeling techniques such as advanced disentangle.

On the other hand, since we develop a very straightforward system using the existing open-source repositories with general settings, there are some possible directions to easily improve our system. First, we adopted only the AIShell3 training set for training our SI models, but adding the AIShell3 testing set into training will increase 30~\% of the training data size, which may provide more diversities of the training data. Second, although we downsampled the target utterances to fit the sampling rate of the AIShell3 training set, simply upsampling the AIShell3 training set to fit the target utterances may achieve higher speech quality because of the higher sampling rate of the synthetic speech. Third, the acoustic features and model hyperparameters were just followed the same setting for generating 24~KHz speech. If these settings are optimized for 44.1~KHz, the performance of our system will be presumably improved.

\section{conclusion}
In this paper, we describe our system submitted to M2VoC 2021. Since our system is straightforward, simple, and easy to reproduce, the system may be suitable as a baseline few-shot TTS system. The discussions of our system provide insights into the performance measurements, the amount of data, and the effectiveness of the speaker adaptation.

%\section{REFERENCES}
%\label{sec:refs}

%List and number all bibliographical references at the end of the
%paper. The references can be numbered in alphabetic order or in
%order of appearance in the document. When referring to them in
%the text, type the corresponding reference number in square
%brackets as shown at the end of this sentence \cite{C2}. An
%additional final page (the fifth page, in most cases) is
%allowed, but must contain only references to the prior
%literature.

\bibliography{reference.bib}
\bibliographystyle{IEEEbib}

%\bibliography{strings,refs}

\end{document}